\documentclass[aps,showpacs,preprintnumbers,amsmath,amssymb,nofootinbibt,preprint]{revtex4}
\usepackage{graphicx}

\begin{document}

%\preprint{gr-qc/yymmnnn}

\title{Modified gravity with arbitrary coupling between matter and geometry}
\author{T. Harko}
\email{harko@hkucc.hku.hk} \affiliation{Department of Physics and
Center for Theoretical and Computational Physics, The University
of Hong Kong, Pok Fu Lam Road, Hong Kong, P. R. China}

\date{\today}

\begin{abstract}
The field equations of a generalized $f(R)$ type gravity model, in
which there is an arbitrary coupling between matter and geometry,
are obtained. The equations of motion for test particles are
derived from a variational principle in the particular case in
which the Lagrange density of the matter is an arbitrary function
of the energy-density of the matter only. Generally, the motion is
non-geodesic, and takes place in the presence of an extra force
orthogonal to the four-velocity. The Newtonian limit of the model
is also considered. The perihelion precession of an elliptical planetary orbit in the presence of an extra force is obtained in a general form, and the magnitude of the extra gravitational effects is constrained in the case of a constant extra force by using Solar System observations.
\end{abstract}

\maketitle

\section{Introduction}

Recent astrophysical observations have provided the astonishing
result that around $95$--$96\%$ of the content of the Universe is
in the form of dark matter $+$ dark energy, with only about
$4$--$5\%$ being represented by baryonic matter \cite{Ri98}. More
intriguingly, around $70\%$ of the energy-density is in the form
of what is called "dark energy", and is responsible for the
acceleration of the distant type Ia supernovae \cite {PeRa03}.
Hence, today's models of astrophysics and cosmology face two
fundamental problems, the dark energy problem, and the dark matter
problem, respectively. Although in recent years many different
suggestions have been proposed to overcome these issues, a
satisfactory answer has yet to be obtained.

A very promising way to explain the observational data is to
assume that at large scales the Einstein gravity model of general
relativity breaks down, and a more general action describes the
gravitational field. Theoretical models in which the standard
Einstein-Hilbert action is replaced by an arbitrary function of
the Ricci scalar $R$, first proposed in \cite{Bu70}, have been
extensively investigated lately. Cosmic acceleration can be
explained by $f(R)$ gravity \cite{Carroll:2003wy}, and the
conditions of viable cosmological models have been derived in
\cite{viablemodels}. In the context of the Solar System regime,
severe weak field constraints seem to rule out most of the models
proposed so far \cite{solartests,Olmo07}, although viable models
do exist \cite
{Hu:2007nk,solartests2,Sawicki:2007tf,Amendola:2007nt}. The
possibility that the galactic dynamic of massive test particles
can be understood without the need for dark matter was also
considered in the framework of $\ f(R)$
gravity models \cite{Cap2,Borowiec:2006qr,Mar1,Boehmer:2007kx,Bohmer:2007fh}%
. For a review of $f(R)$ generalized gravity models see
\cite{SoFa08}.

A generalization of the $f(R)$ gravity theories was proposed in
\cite {Bertolami:2007gv} by including in the theory an explicit
coupling of an arbitrary function of the Ricci scalar $R$ with the
matter Lagrangian density $L_m$. As a result of the coupling the
motion of the massive particles is non-geodesic, and an extra
force, orthogonal to the four-velocity, arises. The connections
with MOND and the Pioneer anomaly were also explored. The
implications of the non-minimal coupling on the stellar
equilibrium were investigated in \cite {Bertolami:2007vu}, where
constraints on the coupling were also obtained. An inequality
which expresses a necessary and sufficient condition to avoid the
Dolgov-Kawasaki instability for the model was derived in
\cite{Fa07}. The relation between the model with geometry-matter
coupling and ordinary scalar-tensor gravity, or scalar-tensor
theories which include non-standard couplings between the scalar
and matter was studied in \cite{SoFa08a}. The motion of matter in
such theories as well as the dark matter problem in galaxies was
also analyzed. In the specific case where both the action and the
coupling are linear in $R$ the action leads to a theory of gravity
which includes higher order derivatives of the matter fields
without introducing more dynamics in the gravity sector
\cite{So08}. The equivalence between a scalar theory and the model
with the non-minimal coupling of the scalar curvature and matter
was considered in \cite{BePa08}. This equivalence allows for the
calculation of the PPN parameters $\beta$ and $\gamma$, which may
lead to a better understanding of the weak-field limit of $f(R)$
theories. Different forms for the matter Lagrangian density ${\cal L}_m$,
and the resulting extra-force, were considered in \cite{BeLoPa08}, and it
was shown that more natural forms for ${\cal L}_m$ do not imply the
vanishing of the extra-force. The impact on the classical equivalence
between different Lagrangian descriptions of a perfect fluid was also analyzed.
Similar couplings between gravitation and matter have
also been considered as possible explanations for the accelerated
expansion of the universe and of the dark energy in \cite{Od}.

In all the previous studies of the models with matter-geometry
coupling, the matter part in the coupling was represented by the
Lagrangian density of the matter, while the geometric part was
considered to be an arbitrary function of the Ricci scalar. We may
call this class of models as modified gravity models with linear
matter-geometry coupling. However, more general models, in which
the matter part in the coupling is an arbitrary function of the
Lagrangian density of the matter, can also be constructed, and
they represent the natural generalization of the models with
linear matter coupling. It is the purpose of this Letter to present
the field equations of a generalized gravity model, in which there
is an arbitrary coupling between matter and geometry. In this
class of models the energy-momentum tensor of the matter is
generally not conserved.  The equations of motion of the test
particles are also obtained, by using a variational principle,  in
the particular case in which the Lagrange function of the matter
is a function of the density of the matter only. In this model
the motion is non-geodesic. The study of the Newtonian limit shows
that the matter-geometry coupling induces a supplementary
acceleration of the test particles. This acceleration may be
responsible for the constancy of the galactic rotation curves,
which is usually attributed to the presence of the dark matter. As an observational test of the model we consider the perihelion precession of an elliptic planetary orbit in the presence of the extra force. The precession angle is derived in a general form, and from the observed value of the perihelion precession of the planet Mercury the magnitude of the extra acceleration is constrained.

The present Letter is organized as follows. The field equations of
the model are derived in Section II. The equations of motion of
the test particles and their Newtonian limit are considered in
Section III for a particular model in which the matter Lagrangian is a function of the density only. We discuss and conclude our results in Section IV.

\section{Gravitational field equations in $f(R)$ type models with arbitrary
coupling between matter and geometry}

The most general action for a $f(R)$ type modified gravity
involving an arbitrary coupling between matter and geometry is
given, in a system of units with $8\pi G=c=1$, by
\begin{equation}
S=\int \left[ \frac{1}{2}f_{1}(R)+G\left( L_{m}\right) f_{2}\left( R\right) %
\right] \sqrt{-g}d^{4}x,
\end{equation}
where $f_{i}(R)$, $i=1,2$ are arbitrary functions of the Ricci
scalar $R$, while $G\left( L_{m}\right) $ is an arbitrary function
of the matter
Lagrangian density $L_{m}$. The only requirement for the functions $f_{i}$, $%
i=1,2$ and $G$ is to be analytical function of the Ricci scalar $R$ and $%
L_{m}$, respectively, that is, they must possess a Taylor series
expansion about any point. When $f_{1}(R)=R$, $f_{2}(R)=1$ and
$G\left( L_{m}\right)
=L_{m}$, we recover standard general relativity. With $f_{2}(R)=1$ and $%
G\left( L_{m}\right) =L_{m}$ we obtain the $f(R)$ generalized
gravity
models. The case $G\left( L_{m}\right) =1+\lambda L_{m}$, $\lambda =\mathrm{%
constant}$, corresponds to the (linear) coupling between matter
and geometry, considered in \cite{Bertolami:2007gv} -
\cite{BeLoPa08}.

We define the energy-momentum tensor of the matter as
\begin{equation}
T_{\mu \nu }=-\frac{2}{\sqrt{-g}}\frac{\delta \left( \sqrt{-g}L_{m}\right) }{%
\delta g^{\mu \nu }}.
\end{equation}

By assuming that the Lagrangian density $L_m$ of the matter
depends only on the metric tensor components, and not on its
derivatives, we obtain $T_{\mu \nu }=L_{m}g_{\mu \nu }-2\partial
L_{m}/\partial g^{\mu \nu }$.

Varying the action with respect to the metric tensor $g_{\mu \nu
}$ we obtain the field equations of the model as
\begin{eqnarray}
&&F_{1}(R)R_{\mu \nu }-\frac{1}{2}f_{1}(R)g_{\mu \nu }+\left(
g_{\mu \nu }\square -\nabla _{\mu }\nabla _{\nu }\right)
F_{1}(R)=-2G\left(
L_{m}\right) F_{2}(R)R_{\mu \nu }  \nonumber  \label{feq} \\
&&-2\left( g_{\mu \nu }\square -\nabla _{\mu }\nabla _{\nu
}\right) G\left(
L_{m}\right) F_{2}(R)-  \nonumber \\
&&f_{2}(R)\left[ K\left( L_{m}\right) L_{m}-G\left( L_{m}\right)
\right] g_{\mu \nu }+f_{2}(R)K\left( L_{m}\right) T_{\mu \nu },
\end{eqnarray}
where we denoted $F_{i}(R)=df_{i}(R)/dR$, $i=1,2$ and $K\left(
L_{m}\right) =dG\left( L_{m}\right) /dL_{m}$, respectively. For
$G\left( L_{m}\right)
=L_{m}$ and by rescaling the function $f_{2}\left( R\right) $ so that $%
f_{2}(R)\rightarrow 1+\lambda f_{2}(R)$, we reobtain the field
equations proposed in \cite{Bertolami:2007gv}.

By contracting the field equations given by Eq.~(\ref{feq}) we
obtain the scalar equation
\begin{eqnarray}
&&3\square \left[ F_{1}(R)+2G\left( L_{m}\right) F_{2}(R)\right]
+\left[
F_{1}(R)+2G\left( L_{m}\right) F_{2}(R)\right] R-  \nonumber \\
&&2f_{1}(R)+4f_{2}(R)\left[ K\left( L_{m}\right) L_{m}-G\left( L_{m}\right) %
\right] =K\left( L_{m}\right) f_{2}(R)T,
\end{eqnarray}
where $T=T_{\mu }^{\mu }$. By taking the covariant divergence of
Eq.(\ref {feq}), with the use of the mathematical identity $\nabla
^{\mu }\left[ a^{\prime }(R)R_{\mu \nu }-a(R)g_{\mu \nu }/2+\left(
g_{\mu \nu }\square -\nabla _{\mu }\nabla _{\nu }\right)
a(R)\right] \equiv 0$ \cite{Ko06}, where $a(R)$ is an arbitrary
function of the Ricci scalar $R$ and $a^{\prime }(R)=da/dR$, we
obtain
\begin{equation}
\nabla ^{\mu }T_{\mu \nu }=\nabla ^{\mu }\ln \left[
f_{2}(R)K\left( L_{m}\right) \right] \left\{ L_{m}g_{\mu \nu
}-T_{\mu \nu }\right\} =2\nabla
^{\mu }\ln \left[ f_{2}(R)K\left( L_{m}\right) \right] \frac{\partial L_{m}}{%
\partial g^{\mu \nu }}.  \label{noncons}
\end{equation}

The requirement of the conservation of the energy-momentum tensor
of matter, $\nabla ^{\mu }T_{\mu \nu }=0$, gives an effective
functional relation between the matter Lagrangian density and the functions $f_{2}(R)$ and $%
K\left( L_{m}\right) $,
\begin{equation}
\nabla ^{\mu }\ln \left[ f_{2}(R)K\left( L_{m}\right) \right]
\frac{\partial L_{m}}{\partial g^{\mu \nu }}=0.
\end{equation}

Thus, once the matter Lagrangian density is known, by an
appropriate choice of the functions $G\left( L_{m}\right) $ and
$f_{2}(R)$ one can construct, at least in principle, conservative
models with arbitrary matter-geometry coupling.

\section{Models with arbitrary density-dependent matter Lagrangian}

As a specific case of generalized gravity models with arbitrary
matter-geometry coupling, we consider the case in which the matter
Lagrangian density is an arbitrary function of the energy density
of the matter $\rho $ only, so that $L_{m}=L_{m}\left( \rho
\right) $. Then the energy-momentum tensor of the matter is given
by
\begin{equation}\label{tens}
T^{\mu \nu }=\rho \frac{dL_{m}}{d\rho }u^{\mu }u^{\nu }+\left(
L_{m}-\rho \frac{dL_{m}}{d\rho }\right) g^{\mu \nu },
\end{equation}
where the four-velocity $u^{\mu }=dx^{\mu }/ds$ satisfies the condition $%
g^{\mu \nu }u_{\mu }u_{\nu }=1$, and we have also used the
relation $\delta \rho =\left( 1/2\right) \rho \left( g_{\mu \nu
}-u_{\mu }u_{\nu }\right) \delta g^{\mu \nu }$.

The energy-momentum tensor given by Eq.~(\ref{tens}) can be written in a
form similar to the perfect fluid case if we assume that the thermodynamic
pressure $p$ obeys a barotropic equation of state, so that $p=p\left( \rho
\right) $. The perfect-fluid type representation can be  obtained by
assuming that the matter Lagrangian satisfies the equations
\begin{equation}
\rho \frac{dL_{m}\left( \rho \right) }{d\rho }=\rho +\rho \Pi \left( \rho
\right) +p\left( \rho \right) ,  \label{eq1}
\end{equation}%
and
\begin{equation}
\rho \frac{dL_{m}\left( \rho \right) }{d\rho }-L_{m}\left( \rho \right)
=p\left( \rho \right) ,  \label{eq2}
\end{equation}
respectively, where $\Pi \left( \rho \right) $ is an arbitrary function of
the density. Substituting the term $\rho dL_{m}\left( \rho \right) /d\rho $
from Eq. (\ref{eq1}) into Eq. (\ref{eq2}) gives the matter Lagrangian as $%
L_{m}\left( \rho \right) =\rho +\rho \Pi \left( \rho \right) $. With this
form of $L_{m}$, Eq. (\ref{eq1}) gives the following differential equation
for $\Pi \left( \rho \right) $,
\begin{equation}
\rho ^{2}\frac{d\Pi \left( \rho \right) }{d\rho }=p\left( \rho \right) ,
\end{equation}%
with the general solution given by
\begin{equation}
\Pi \left( \rho \right) =\int_{0}^{\rho }\frac{p}{\rho ^{2}}d\rho
=\int_{0}^{p}\frac{dp}{\rho }-\frac{p\left(\rho \right)}{\rho }.
\end{equation}

Therefore the matter Lagrangian and the energy-momentum tensor can be
written as
\begin{equation}
L_{m}\left( \rho \right) =\rho \left(1+\int_{0}^{p}\frac{dp}{\rho }\right)-p\left( \rho \right),
\end{equation}
and
\begin{equation}\label{tens1}
T^{\mu \nu }=\left[ \rho +p\left( \rho \right) +\rho \Pi \left( \rho \right) %
\right] u^{\mu }u^{\nu }-p\left( \rho \right) g^{\mu \nu },
\end{equation}
respectively. From a physical point of view $\Pi \left(\rho \right)$ can be interpreted as the elastic (deformation) potential energy of the body, and therefore Eq.~(\ref{tens1}) corresponds to the energy-momentum tensor of a compressible elastic isotropic system.

By imposing the
condition of the conservation of the matter current, $\nabla _{\nu
}\left( \rho u^{\nu }\right) =0$, and with the use of the identity
$u^{\nu }\nabla _{\nu }u^{\mu }=d^{2}x^{\mu }/ds^{2}+\Gamma _{\nu
\lambda }^{\mu }u^{\nu }u^{\lambda }$, from Eq.~(\ref {noncons})
we obtain the equation of motion of a test particle in the
modified gravity model as
\begin{equation}
\frac{d^{2}x^{\mu }}{ds^{2}}+\Gamma _{\nu \lambda }^{\mu }u^{\nu
}u^{\lambda }=f^{\mu },  \label{eqmot}
\end{equation}
where
\begin{equation}
f^{\mu }=-\nabla _{\nu }\ln \left\{ f_{2}(R)K\left[ L_{m}\left( \rho \right) %
\right] \frac{dL_{m}\left( \rho \right) }{d\rho }\right\} \left(
u^{\mu }u^{\nu }-g^{\mu \nu }\right) .
\end{equation}

The extra-force $f^{\mu }$, generated due to the presence of the
coupling between matter and geometry, is perpendicular to the
four-velocity, $f^{\mu }u_{\mu }=0$. The equation of motion Eq.~
(\ref{eqmot}) can be obtained from the variational principle
\begin{equation}
\delta S_{p}=\delta \int L_{p}ds=\delta \int \sqrt{Q}\sqrt{g_{\mu
\nu }u^{\mu }u^{\nu }}ds=0,  \label{actpart}
\end{equation}
where $S_{p}$ and $L_{p}=\sqrt{Q}\sqrt{g_{\mu \nu }u^{\mu }u^{\nu
}}$ are the action and the Lagrangian density for the test
particles, respectively, and
\begin{equation}
\sqrt{Q}=f_{2}(R)K\left[ L_{m}\left( \rho \right) \right]
\frac{dL_{m}\left( \rho \right) }{d\rho }.  \label{Q}
\end{equation}

To prove this result we start with the Lagrange equations
corresponding to the action~(\ref{actpart}),
\begin{equation}
\frac{d}{ds}\left( \frac{\partial L_{p}}{\partial u^{\lambda }}\right) -%
\frac{\partial L_{p}}{\partial x^{\lambda }}=0.
\end{equation}

Since $\partial L_{p}/\partial u^{\lambda }=\sqrt{Q}u_{\lambda }$ and $%
\partial L_{p}/\partial x^{\lambda }=\left( 1/2\right) \sqrt{Q}g_{\mu \nu
,\lambda }u^{\mu }u^{\nu }+\left( 1/2\right) Q_{,\lambda }/Q$, a
straightforward calculation gives the equations of motion of the
particle as
\begin{equation}
\frac{d^{2}x^{\mu }}{ds^{2}}+\Gamma _{\nu \lambda }^{\mu }u^{\nu
}u^{\lambda }+\left( u^{\mu }u^{\nu }-g^{\mu \nu }\right) \nabla
_{\nu }\ln \sqrt{Q}=0.
\end{equation}
By simple identification with the equation of motion of the
modified gravity model with arbitrary matter-geometry coupling,
given by Eq.~(\ref{eqmot}), we obtain the explicit form of
$\sqrt{Q}$ as given by Eq.~(\ref{Q}).

The variational principle~(\ref{actpart}) can be used to study the
Newtonian limit of the model. In the limit of weak gravitational
fields, $ds\approx
\sqrt{1+2\phi -\vec{v}^{2}}dt\approx \left( 1+\phi -\vec{v}^{2}/2\right) dt$%
, where $\phi $ is the Newtonian potential and $\vec{v}$ is the
usual tridimensional velocity of the particle. By representing the function $\sqrt{%
Q}$ as
\begin{equation}
\sqrt{Q}=f_{2}(R)K\left[ L_{m}\left( \rho \right) \right]
\frac{dL_{m}\left(
\rho \right) }{d\rho }=1+U\left( R,L_{m}\left( \rho \right) ,\frac{%
dL_{m}\left( \rho \right) }{d\rho }\right) ,
\end{equation}
where $U<<1$, the equations of motion of the particle can be
obtained from the variational principle
\begin{equation}
\delta \int \left[ U\left( R,L_{m}\left( \rho \right)
,\frac{dL_{m}\left( \rho \right) }{d\rho }\right) +\phi
-\frac{\vec{v}^{2}}{2}\right] dt=0,
\end{equation}
and are given by
\begin{equation}
\vec{a}=-\nabla \phi -\nabla U=\vec{a}_{N}+\vec{a}_{E},
\end{equation}
where $\vec{a}_{N}=-\nabla \phi $ is the usual Newtonian
gravitational acceleration, and $\vec{a}_{E}=-\nabla U$ is a
supplementary acceleration induced due to the coupling between
matter and geometry.

An estimation of the effect of the extra-force, generated by the coupling
between matter and geometry, on the orbital parameters of the motion of the
planets around the Sun can be obtained in a simple way by using the
properties of the Runge-Lenz vector, defined as $\vec{A}=\vec{v}$ $\times
\vec{L}-\alpha \vec{e}_{r}$, where $\vec{v}$ is the velocity relative to the
Sun, with mass $M_{\odot}$,  of a planet of mass $m$, $\vec{r}=r\vec{e}_{r}$
is the two-body position vector, $\vec{p}=\mu \vec{v}$ is the relative
momentum,  $\mu =mM_{\odot }/\left( m+M_{\odot}\right) $ is the reduced mass, $\
\ \vec{L}=\vec{r}$ $\times \vec{p}=\mu r^{2}\dot{\theta}\vec{k}$ is the
angular momentum, and $\alpha =GmM_{\odot}$ \cite{prec}. For an elliptical orbit of
eccentricity $e$, major semi-axis $a$, and period $T$, the equation of the orbit is given by $\left( L^{2}/\mu
\alpha \right) r^{-1}=1+e\cos \theta $.  The Runge-Lenz vector can
be expressed as $\vec{A}=\left( \vec{L}^{2}/\mu r-\alpha \right) \vec{e}_{r}-%
\dot{r}L\vec{e}_{\theta }$, and its  derivative with respect to the polar
angle $\theta $ is given by $d\vec{A}/d\theta =r^{2}\left[ dV(r)/dr-\alpha
/r^{2}\right] \vec{e}_{\theta }$, where $%
V(r)$ is the potential of the central force \cite{prec}. The potential term consists of the
Post-Newtonian potential, $V_{PN}(r)=-\alpha /r-3\alpha ^{2}/mr^{2}$,
plus the contribution from the general coupling between matter and geometry.
Thus we have $d\vec{A}/d\theta =r^{2}\left[ 6\alpha ^{2}/mr^{3}+m\vec{a}%
_{E}(r)\right] \vec{e}_{\theta }$, where we have also assumed that $\mu \approx m$. The change in direction $\Delta \phi $ of
the perihelion with a change of $\theta $ of $2\pi $ is obtained as $\Delta
\phi =\left( 1/\alpha e\right) \int_{0}^{2\pi }\left\vert \dot{\vec{L}}\times d%
\vec{A}/d\theta \right\vert d\theta $, and it is given by
\begin{equation}\label{prec}
\Delta \phi =24\pi ^{3}\left( \frac{a}{T}\right) ^{2}\frac{1}{1-e^{2}}+\frac{%
L}{8\pi ^{3}me}\frac{\left( 1-e^{2}\right) ^{3/2}}{\left( a/T\right) ^{3}}%
\int_{0}^{2\pi }\frac{a_{E}\left[ L^{2}\left( 1+e\cos \theta \right)
^{-1}/m\alpha \right] }{\left( 1+e\cos \theta \right) ^{2}}\cos \theta
d\theta ,
\end{equation}%
where we have used the relation $\alpha /L=2\pi \left( a/T\right) /\sqrt{%
1-e^{2}}$. The first term of this equation corresponds to the standard general relativistic precession of the perihelion of the planets, while the second term gives the contribution to the perihelion precession due to the presence of the coupling between matter and geometry.

As an example of the application  of Eq.~(\ref{prec}) we consider the case for which the extra-force may be considered as a constant,  $a_E\approx$ constant, an approximation that could be valid for small regions of the space-time.
%MOND type acceleration $a_E\approx \sqrt{a_0a_N}=\sqrt{GM_{\odot}a_0}/r$, where $a_0$ is a constant acceleration, which was proposed phenomenologically as a dynamical model for dark matter \cite{Milgrom}. In the Newtonian limit the extra-acceleration generated by the coupling between matter and geometry can be expressed in a similar form \cite{Bertolami:2007gv}.
  With the use of Eq.~(\ref{prec}) one finds for the perihelion precession
the expression
\begin{equation}\label{prec1}
\Delta \phi =\frac{6\pi GM_{\odot}}{a\left( 1-e^{2}\right) }+\frac{2\pi a^{2}%
\sqrt{1-e^{2}}}{GM_{\odot}}a_{E},
\end{equation}
where we have also used Kepler's third law, $T^2=4\pi ^2a^3/GM_{\odot}$. For the planet Mercury $a=57.91\times 10^{11}$ cm, and $e=0.205615$,
respectively, while $M_{\odot }=1.989\times 10^{33}$ g.  With these
numerical values the first term in Eq. (\ref{prec1}) gives the standard
general relativistic value for the precession angle, $\left( \Delta \phi
\right) _{GR}=42.962$ arcsec per century, while the observed value of the precession is $\left(\Delta \phi \right)_{obs}=43.11\pm0.21$ arcsec per century \cite{merc}. Therefore the difference $\left(\Delta \phi \right)_{E}=\left(\Delta \phi \right)_{obs}-\left( \Delta \phi
\right) _{GR}=0.17$ arcsec per century can be attributed to other physical effects. Hence the observational constraints requires
that the value of the constant $a_E$ must satisfy the condition $a_E\leq 1.28\times 10^{-9}$ cm/s$^2$. This value of $a_E$, obtained from the
solar system observations, is somewhat smaller than the value of the extra-acceleration $%
a_{0}\approx 10^{-8}$ cm/s$^{2}$, necessary to explain
the "dark matter" properties, as well as the Pioneer anomaly \citep{Bertolami:2007gv}. However, it does not rule out the possibility of the presence of some extra gravitational effects acting at both the solar system and galactic levels, since the assumption of a constant extra-force is not correct on larger astronomical scales.

\section{Conclusions}

In the present Letter we have considered a generalized gravity
model with an arbitrary coupling between matter and geometry,
described by the product of an arbitrary function of the Lagrange
density of the matter, and an arbitrary function of the Ricci
scalar. The proposed action represents the most general extension
of the standard Hilbert action for the gravitational field,
$S=\int \left[ R/2+L_{m}\right] \sqrt{-g}d^{4}x$. The equations of
motion corresponding to this model show the presence of an
extra-force acting on test particles, and the motion is generally
non-geodesic. The physical implications of such a force have been
already analyzed in the framework of the generalized gravity model
with linear coupling between matter and geometry, considered in
\cite{Bertolami:2007gv}, and the possible implications for the
dark matter problem and for the explanation of the Pioneer anomaly
have also been investigated. On the other hand, the field equations
Eqs.~(\ref{feq}) are equivalent to the Einstein equations of the
$f(R)$ model in empty space-time, but differ from them, as well as
from standard general relativity, in the presence of matter.
Therefore the predictions of the present model could lead to some
major differences, as compared to the predictions of standard
general relativity, in several problems of current interest, like
cosmology, gravitational collapse or the generation of
gravitational waves. The study of these phenomena may also provide
some specific signatures and effects, which could distinguish and
discriminate between the various gravity models.

\section*{Acknowledgments}

I would like to thank to the anonymous referee, whose comments and suggestions helped me to significantly improve the manuscript.
This work is supported by an RGC grant of the government of the
Hong Kong SAR.

\end{document}